\documentclass[%
 reprint,
 superscriptaddress,
 amsmath,amssymb,
]{revtex4-1}

\usepackage{graphicx}
\usepackage{dcolumn}
\usepackage{bm}
\usepackage{hyperref}
\usepackage{url}
\usepackage{comment}
\usepackage{xspace}
\usepackage{mathtools}
\usepackage{framed}
\usepackage{xcolor}
\usepackage{acro}

\colorlet{shadecolor}{orange!15}
\renewenvironment{quote}{\begin{shaded*}\begin{oldquote}}{\end{oldquote}\end{shaded*}}

\newcommand{\sref}[1]{Sec.~(\ref{#1})}
\newcommand{\Sref}[1]{Section~(\ref{#1})}
\newcommand{\fref}[1]{Fig.~\ref{#1}}
\newcommand{\Fref}[1]{Figure~\ref{#1}}

\newcommand{\appropto}{\mathrel{\vcenter{
  \offinterlineskip\halign{\hfil$##$\cr
    \propto\cr\noalign{\kern2pt}\sim\cr\noalign{\kern-2pt}}}}}

\newcommand\Ne{$n_\mathrm{e}$\xspace}
\newcommand\Te{$T_\mathrm{e}$\xspace}

\newcommand\Wdia{$W_\mathrm{dia}$\xspace}

\newcommand{\mycomment}[1]{}
\excludecomment{derivation}

\begin{document}
\newcommand{\ManuscriptTitle}{
    Pellet-Size Scaling of Quasi-Steady-State Plasma Performance in Wendelstein 7-X
}

\title{\ManuscriptTitle}

\author{Keisuke Fujii}
\email{fujiik@ornl.gov}
\affiliation{%
    Fusion Energy Division, Oak Ridge National Laboratory, Oak Ridge, TN 37831-6305, United States of America
}
\author{Edgardo Villalobos Granados} 
\author{Maryam Huck}
\author{J\"urgen Baldzuhn}
\affiliation{%
    Max Planck Institute for Plasma Physics, Wendelsteinstrasse 1, 17491 Greifswald, Germany
}
\author{Naoki Tamura}
\affiliation{%
    Max Planck Institute for Plasma Physics, Wendelsteinstrasse 1, 17491 Greifswald, Germany
}
\author{Steven Meitner}
\affiliation{%
    Fusion Energy Division, Oak Ridge National Laboratory, Oak Ridge, TN 37831-6305, United States of America
}
\author{Larry Baylor}
\affiliation{%
    Fusion Energy Division, Oak Ridge National Laboratory, Oak Ridge, TN 37831-6305, United States of America
}
\affiliation{%
    Type One Energy, 2410 Cherahala Blvd., Knoxville, TN 37931, United States of America
}
\author{Golo Fuchert}
\author{Kai Jakob Brunner}
\author{Jens Knauer}
\author{Ekkehard Pasch}
\author{Jannik Wagner}
\affiliation{%
    Max Planck Institute for Plasma Physics, Wendelsteinstrasse 1, 17491 Greifswald, Germany
}
\author{Bart Lomanowski}
\affiliation{%
    Fusion Energy Division, Oak Ridge National Laboratory, Oak Ridge, TN 37831-6305, United States of America
}
\author{W7-X Team}
\thanks{See O.~Grulke et al., Nucl.\ Fusion \textbf{66}, 116003 (2026), \href{https://doi.org/10.1088/1741-4326/ae5f32}{doi:10.1088/1741-4326/ae5f32}, for the full list of W7-X Team members.}
\affiliation{%
    Max Planck Institute for Plasma Physics, Wendelsteinstrasse 1, 17491 Greifswald, Germany
}

\date{\today}

\begin{abstract}
    A continuous cryogenic-pellet injector has been used to realize long-pulse high-performance plasmas in Wendelstein 7-X (W7-X).
    A hydrogen ice pellet deposits particles directly in the confined region. If the particles are deposited sufficiently far inside the plasma, they produce a density gradient that suppresses ion-temperature-gradient turbulence and improves plasma confinement.
    However, after multiple pellets have been injected and the plasma density and temperature have increased, the plasma performance begins to saturate.
    In this work, we analyze multiple pellet-injection experiments conducted in 2024 and 2025, during which the injected pellet sizes varied unintentionally.
    This analysis reveals a positive correlation between pellet size and the quasi-steady-state stored energy of W7-X plasmas.
    %
    %
    Although this pellet-size dependence can be understood qualitatively from pellet-ablation physics, the measured deposition position differs quantitatively from the neutral-gas-shielding (NGS) model prediction. This discrepancy suggests significant inward transport of the pellet cloud.
    The trend identified here suggests that injection of even larger pellets could further improve plasma performance.
\end{abstract}

\maketitle

\section{Introduction}

\begin{figure*}[t]
    \centering
    \includegraphics[width=\textwidth]{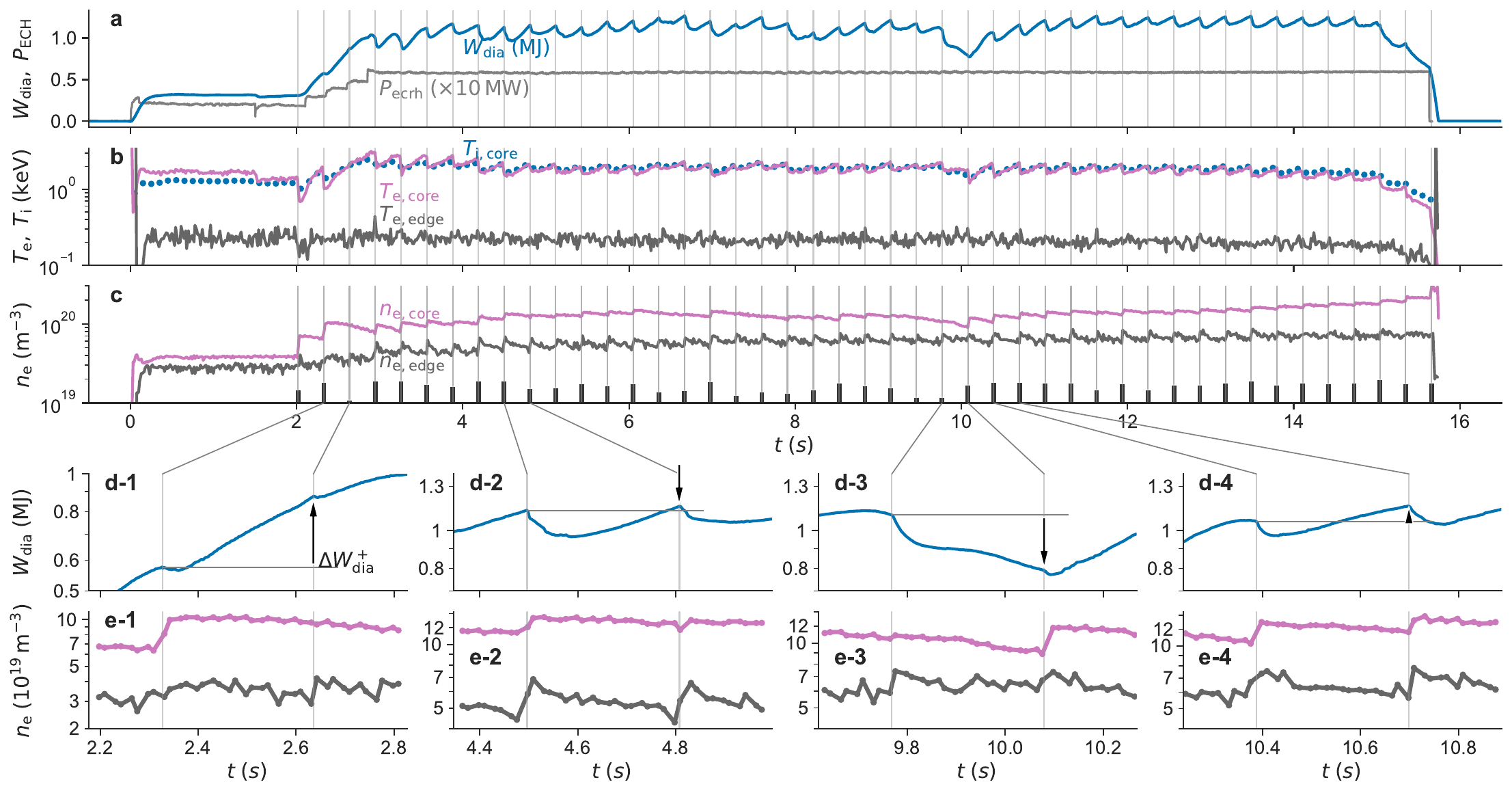}
    \caption{
        Shot summary for a typical W7-X experiment with continuous hydrogen pellet injection (program ID: 20250327.031).
        (a) Temporal evolution of the plasma stored energy $W_\mathrm{dia}$ and heating power $P_\mathrm{ecrh}$.
        (b) Core ion temperature $T_\mathrm{i,core}$, the core electron temperature $T_\mathrm{e,core}$, and edge electron temperature $T_\mathrm{e,edge}$.
        (c) Core and edge electron density ($n_\mathrm{e,core}$ and $n_\mathrm{e,edge}$, respectively).
        The vertical lines indicate the pellet-injection times.
        The vertical thick bars in (c) indicate the pellet size for each injection. The pellet size was varied for each injection event.
        (d, e) Expanded views of $W_\mathrm{dia}$ and $n_\mathrm{e}$ for selected pellet-injection events.
        Depending on the plasma state and the pellet size, the plasma response is different. 
        Panels (d-1, e-1) show an injection event while the plasma stored energy is still low ($\approx 0.5$ MJ).
        Panels (d-2, e-2)--(d-4, e-4) illustrate the responses of higher-energy plasmas to pellets of different sizes.
        The definition of $\Delta W_\mathrm{dia}^+$ is also shown in (d-1).
    }
    \label{fig:summary}
\end{figure*}

\begin{figure}[t]
    \centering
    \includegraphics[width=\columnwidth]{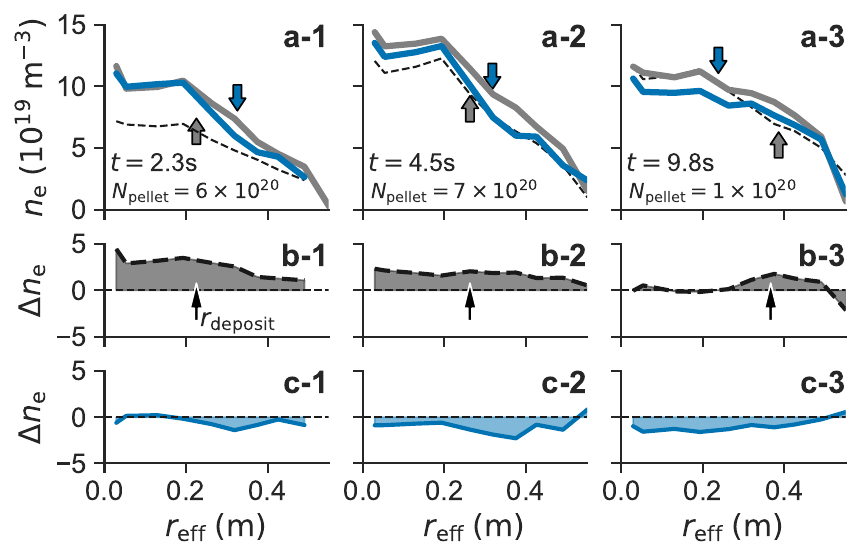}
    \caption{
        Profile evolution of $n_\mathrm{e}$ for typical pellet injection events.
        Columns 1, 2, and 3 correspond to the events shown in (d-1, e-1), (d-2, e-2), and (d-3, e-3), respectively, in \fref{fig:summary}.
        The dotted line shows the profile just before the pellet injection event, while the thick gray line shows the profile just after the event.
        The blue curve shows the \Ne profile 0.2~s after the injection, roughly when $W_\mathrm{dia}$ reaches its maximum.
        The second row shows the density increase caused by pellet injection (the difference between the thick gray and dotted lines in the top panels).
        The third row shows the subsequent density change (the difference between the blue and thick gray lines in the top panels).
    }
    \label{fig:profile}
\end{figure}

Efficient core particle fueling is a central requirement for reactor-relevant magnetically confined plasmas. 
In large fusion devices, gas puffing predominantly supplies particles near the plasma boundary, and the resulting edge-localized source can be insufficient for producing and sustaining a peaked density profile in the core. 
Cryogenic hydrogen pellet injection provides a more direct actuator for core fueling because a solid pellet can cross flux surfaces before being ablated by the background plasma \cite{Pegourie2007,Parks1978,Milora1978}.

In the optimized stellarator Wendelstein 7-X (W7-X), peaked density profiles are particularly important for achieving high-performance plasmas.
In the W7-X magnetic configuration, density-gradient-driven trapped-electron modes are expected to be strongly reduced by the quasi-isodynamic, maximum-$J$ geometry~\cite{Proll2012-prl,Xanthopoulos2014-arxiv}.
Ion-temperature-gradient (ITG) turbulence therefore plays a central role in the remaining turbulent transport~\cite{Alcuson2020-pl,Stechow2020-arxiv,Carralero2021,Ford2024}.

Oak Ridge National Laboratory delivered a continuous pellet fueling system (CPFS) to W7-X, where it has been used for core fueling~\cite{Meitner2020-aa}.
Once injected, a hydrogen pellet penetrates the plasma and is ablated by the incident plasma heat flux, providing a localized particle source.
This pellet-based core fueling generates a density gradient and a peaked profile~\cite{Bozhenkov2020-hl}.
Recent W7-X experiments have therefore used pellet injection as an essential tool to access high-density-gradient, high-performance scenarios.

The CPFS has also been essential for sustaining high performance in long-pulse plasmas.
Repeated pellet injection can restore a high-performance plasma state, thereby enabling quasi-steady-state operation.
This capability contributed to achieving the highest triple product among fusion-device discharges lasting at least 40~s~\cite{Grulke2026-we}.
However, the pellet-based performance improvement does not continue indefinitely; the improvement saturates after multiple injections.

During OP2.2 (2024) and OP2.3 (2025), the CPFS pellet size varied substantially and unintentionally.
Taking advantage of this unplanned variability in pellet size, we study the effect of pellet size on steady-state plasma performance in W7-X.
For this purpose, we collect the pellet-injection events from these campaigns and statistically characterize the plasma response.

The remainder of the paper is organized as follows.
In \sref{sec:typical},
we present a typical pellet-injection experiment in W7-X and discuss how pellet injection improves and sustains a high-performance plasma.
The dependencies on the plasma state and pellet size are already visible in this experiment.
\Sref{sec:analysis} describes our statistical analysis of pellet-injection experiments in OP2.2 and OP2.3, and \sref{sec:discussion} compares the measured deposition depth with the NGS model.
Finally, \sref{sec:summary} summarizes the main conclusions.

\section{Pellet Fueling Experiment in W7-X\label{sec:typical}}

\Fref{fig:summary} shows the shot summary of a typical pellet-injection experiment in W7-X (program ID: 20250327.031).
The plasma is sustained by electron cyclotron resonance heating (ECRH; approximately 5~MW at $t > 3$~s).
\Fref{fig:summary}~(a) shows the temporal evolution of the plasma stored energy \Wdia, measured by the W7-X diamagnetic-loop system~\cite{Rahbarnia2018-dia}, which we use as a proxy for confinement performance.
The pellet-injection times are shown by thin vertical lines. Injection begins at $t=2.00$~s and continues at 3~Hz until $t=15.75$~s.
One of the key diagnostics used in this paper is Thomson scattering~\cite{Bozhenkov2017-ts,Damm2019-ts}.
This system measures the spatial profiles of \Te (lines in \fref{fig:summary}~(b)) and \Ne (\fref{fig:summary}~(c)) at a sampling frequency of approximately 100~Hz.
The ion temperature profile is measured by an X-ray imaging crystal spectrometer~\cite{Kwak2021-bayes} (\fref{fig:summary}~(b) markers).
Here, \textit{core} denotes an average over $r_\mathrm{eff}=0.1$--$0.2$~m, and \textit{edge} denotes an average over $r_\mathrm{eff}=0.4$--$0.5$~m.
The effective minor radius of the last closed flux surface is approximately 0.60~m.
Because \fref{fig:summary}~(b) and (c) use logarithmic vertical axes, the spacing between the core and edge curves is proportional to the inverse density scale length.
We define $L_n^{-1}\equiv-d(\ln n_\mathrm{e})/dr$ and approximate it using the \textit{core} and \textit{edge} values as $L_n^{-1}\equiv[\ln(n_\mathrm{e,core})-\ln(n_\mathrm{e,edge})]/(r_\mathrm{edge}-r_\mathrm{core})$.
Typical radial profiles of \Ne are shown in \fref{fig:profile}.

The thick vertical bars in \fref{fig:summary}~(c) indicate the pellet size for each injection event.
Pellet size is estimated from the signal of a microwave cavity installed along the pellet path in the CPFS.
Although the density axis is logarithmic, the bar heights represent pellet size on a linear scale.

During the initial phase of the experiment (before the pellet injections, $t<2$~s), the plasma density gradient is nearly zero; that is, the core and edge values of \Ne are almost identical, and \Wdia remains below approximately 0.4~MJ.
After the first few pellet injections ($t \lesssim 3$~s), together with the increase in heating power, \Wdia rises to approximately 1~MJ.
Expanded views are provided in \fref{fig:summary}~(d-1) and (e-1).
$n_\mathrm{e,core}$ increases sharply while $n_\mathrm{e,edge}$ remains nearly unchanged, resulting in a steep density gradient.
Once this gradient is established, \Wdia increases significantly.

After approximately the fourth pellet injection, both \Wdia and $L_n^{-1}$ (the spacing between $n_\mathrm{e,core}$ and $n_\mathrm{e,edge}$) begin to saturate.
A typical example is shown in \fref{fig:summary}~(d-2) and (e-2).
The value of \Wdia drops quickly after pellet injection and gradually recovers to its previous value.
The behavior of \Ne differs markedly from that during the first three injections (compare with (d-1) and (e-1)): the logarithmic increase in $n_\mathrm{e,edge}$ is larger than that in $n_\mathrm{e,core}$, whereas $n_\mathrm{e,edge}$ also decays more rapidly.
Thus, the density gradient initially decreases after injection and then recovers as the edge density decays.
\Wdia behaves consistently with the change in the density gradient.

\Fref{fig:profile}~(a) shows changes in the \Ne profiles during pellet injection and the subsequent decay, as measured by Thomson scattering.
The dotted line shows the \Ne profile before the injection, while the gray bold line shows the profile just after the injection.
The blue lines show the \Ne profiles 0.2~s after injection.
\Fref{fig:profile}~(b) and (c) show the differences in the \Ne profiles (b) before and immediately after injection and (c) immediately and 0.2~s after injection.

For the first example, highlighted in \fref{fig:summary}~(d-1) and (e-1), pellet injection produces a substantial density increase in the core.
0.2~s after the injection, the density at the edge decays and the density gradient further increases.
For the second example (the injection event at $t=4.5$~s; \fref{fig:summary}~(d-2) and (e-2)), particles are deposited closer to the edge. The edge density then decays more rapidly, allowing the density gradient to be sustained.

In this experiment, some injection events involved unusually small pellets because of injector-alignment issues.
A good example is the injection event at $t = 9.8$~s, which is highlighted in \fref{fig:summary}~(d-3) and (e-3), as well as \fref{fig:profile}~(a-3).
When a small pellet is injected, particles are deposited only near the edge, while \Ne decays in both the edge and core regions.
This behavior reduces the density gradient, and \Wdia therefore continues to decrease.

In \fref{fig:profile}~(b-1)--(b-3), the average deposition position $r_\mathrm{deposit}$ is shown by the arrow positions.
This deposition position is calculated as the mean radial position of the pellet-induced density increase.
The injection at $t=2.3$~s produces the innermost deposition, whereas the very small pellet injected at $t=9.8$~s produces the outermost deposition.

The Thomson-scattering measurement location is separated toroidally from the pellet-injection location by approximately $41^\circ$, and the delay $\Delta t$ between injection and measurement varies among events.
We use Thomson-scattering frames acquired within $3~\mathrm{ms}<\Delta t<15~\mathrm{ms}$ after injection.
As discussed in \sref{sec:discussion}, this interval is long relative to the equilibration time but short relative to both the fast inward-transport time ($\sim30$~ms~\cite{Damm2026-lj}) and the diffusive time, $a_0^2/D\sim10^{-1}$~s.
Here, $a_0\approx0.6$~m is the plasma minor radius and $D\sim10^0~\mathrm{m^2/s}$ is the diffusion coefficient.
The effect of variations in measurement timing is therefore expected to be small and is further reduced by averaging over multiple injection events in the statistical analysis.

\section{Statistical Analysis of the Pellet Injections\label{sec:analysis}}

\begin{figure}[t]
    \centering
    \includegraphics[width=\columnwidth]{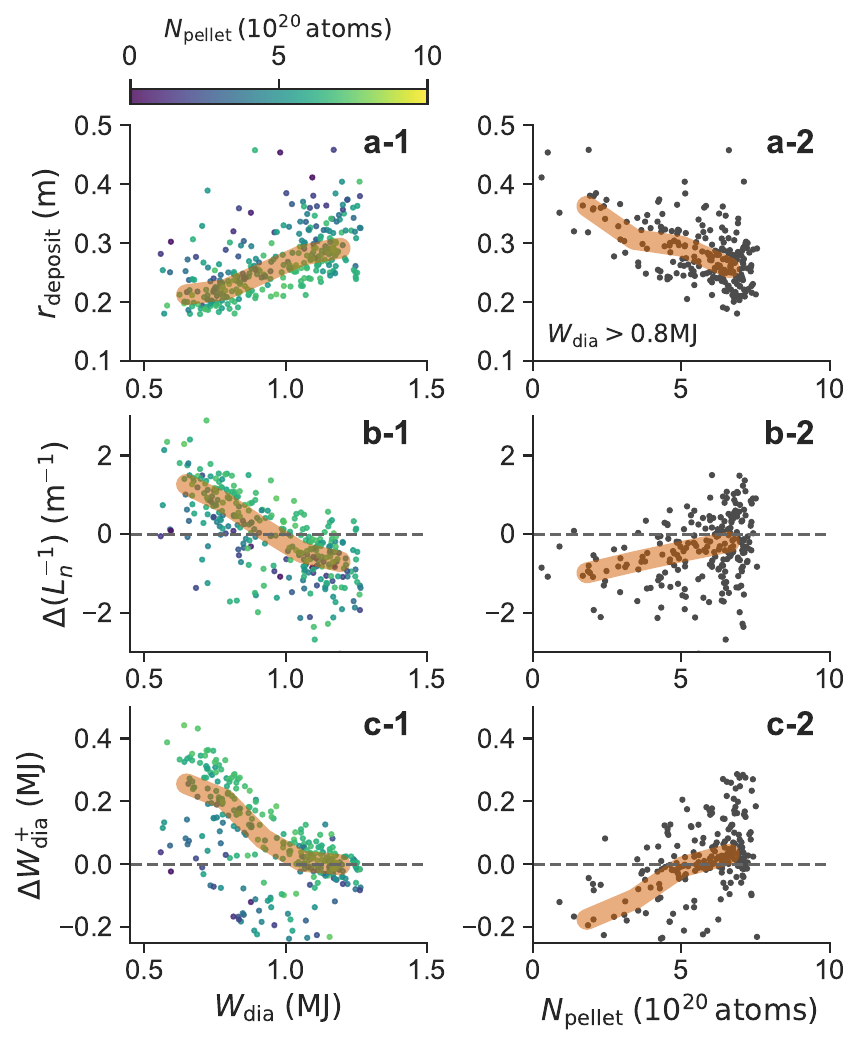}
    \caption{
        Statistics of pellet-injection events throughout OP2.2 and OP2.3.
        Panels (a-1), (b-1), and (c-1) show the dependence of $r_\mathrm{deposit}$, $\Delta L_n^{-1}$, and $\Delta W_\mathrm{dia}^+$, respectively, on $W_\mathrm{dia}$ immediately before pellet injection.
        Here, $\Delta L_n^{-1}$ is the pellet-induced change in the density gradient, measured immediately before and after injection, whereas $\Delta W_\mathrm{dia}^+$ is the net gain in $W_\mathrm{dia}$ after the subsequent profile evolution; its definition is illustrated in \fref{fig:summary}~(d-1).
        The color for each marker shows the size of the injected pellet.
        Panels (a-2), (b-2), and (c-2) show the dependence on pellet size $N_\mathrm{pellet}$ for events with $W_\mathrm{dia}>0.8$~MJ.
        The solid line in each panel shows the average trend.
    }
    \label{fig:statistics}
\end{figure}

As shown in the previous section, the effect of the pellet injection depends both on the plasma state and the pellet size.
To disentangle these effects, we collect the pellet-injection events from OP2.2 and OP2.3 and perform a statistical analysis.

For the analysis presented in this section, we consider only pellet-injection events satisfying the following criteria:
\begin{enumerate}
    \item The plasma heating power $P_\mathrm{ecrh}$ is steady and between 5 and 6~MW.
    \item No neutral-beam injection is applied.
    \item The pellet injection interval is more than 0.3 s.
    \item The magnetic configuration is FTM (high-$\iota$ configuration).
\end{enumerate}
The analysis including all injection events is provided in the Appendix.

\Fref{fig:statistics}~(a-1) shows the \Wdia dependence of $r_\mathrm{deposit}$.
Each marker represents a pellet-injection event, and its color indicates the pellet size ($N_\mathrm{pellet}$; see the color bar at the top).
The horizontal coordinate gives \Wdia immediately before pellet injection.
The orange line overlaid on the markers indicates the overall trend, calculated using averages within narrow bins.
A clear dependence of $r_\mathrm{deposit}$ on \Wdia is observed.
In plasmas with higher stored energy, the deposition position is located farther outward.

\Fref{fig:statistics}~(b-1) shows the pellet-induced change in the density gradient, defined as $\Delta L_n^{-1} \equiv L_{n,\mathrm{after}}^{-1} - L_{n,\mathrm{before}}^{-1}$.
Here, $L_{n,\mathrm{before}}^{-1}$ and $L_{n,\mathrm{after}}^{-1}$ are the logarithmic density gradient before and after the pellet injection, respectively.
$\Delta L_n^{-1}$ decreases with increasing \Wdia and becomes negative for $W_\mathrm{dia} \gtrsim 0.8 \,\mathrm{MJ}$.

\Fref{fig:statistics}~(c-1) shows the eventual increase in \Wdia caused by each injection event (or the decrease when the value is negative).
$\Delta W_\mathrm{dia}^+$ is calculated from the maximum values after each pellet injection event but before the next event.
More precisely, we take the maximum value after the post-injection minimum in \Wdia. The definition of $\Delta W_\mathrm{dia}^+$ is illustrated by arrows in \fref{fig:summary}~(d).
$\Delta W_\mathrm{dia}^+$ shows a clear negative correlation with \Wdia.
Near the maximum value of \Wdia, $\Delta W_\mathrm{dia}^+$ approaches zero, indicating \textit{saturation} of \Wdia.
Pellet injection can increase \Wdia when its initial value is low, but the improvement diminishes as \Wdia increases; without additional heating, \Wdia eventually cannot be increased further.

\Fref{fig:statistics}~(a-2)--(c-2) show the same quantities as functions of $N_\mathrm{pellet}$.
To reduce the influence of \Wdia, we further restrict these plots to events with $W_\mathrm{dia}\geq0.8$~MJ.
The deeper penetration of larger pellets is evident.
Larger pellets also produce a steeper density gradient and a greater increase in \Wdia.

As described in the next section (Sec.~\ref{sec:discussion}), the neutral gas shielding (NGS) model predicts the negative \Ne and \Te dependence and positive $N_\mathrm{pellet}$ dependence of the pellet penetration length.
The trend observed in \fref{fig:statistics}~(a) is qualitatively consistent with this scaling.
A higher-\Wdia plasma generally has higher \Ne and \Te; consequently, the pellet is ablated more quickly and deposits particles at a larger radius.
Conversely, a larger pellet survives longer and therefore penetrates deeper.

The negative dependence of $\Delta L_n^{-1}$ on \Wdia and its positive dependence on $N_\mathrm{pellet}$ may also be understood through this dependence of $r_\mathrm{deposit}$: particles deposited farther inside generate a steeper density gradient.
The subsequent $\Delta W_\mathrm{dia}^+$ may therefore also be influenced by $r_\mathrm{deposit}$.
The steeper density gradient generated by pellet injection increases the plasma stored energy through suppression of ITG turbulence~\cite{Alcuson2020-pl}.

We note that the time scales of the quantities shown in \fref{fig:statistics} differ significantly: \Wdia is evaluated at $\Delta t=0$~s; $r_\mathrm{deposit}$ and $\Delta L_n^{-1}$ are measured at $\Delta t\approx10^{-2}$~s, as determined by the Thomson-scattering timing; and $\Delta W_\mathrm{dia}^+$ is measured at $\Delta t\approx10^{-1}$~s.
The clear trends among quantities spanning these disparate time scales suggest that pellet injection establishes the initial condition from which the plasma subsequently evolves.
These trends also suggest that $r_\mathrm{deposit}$ significantly influences the subsequent plasma performance.

\begin{figure}[h]
    \centering
    \includegraphics[width=\columnwidth]{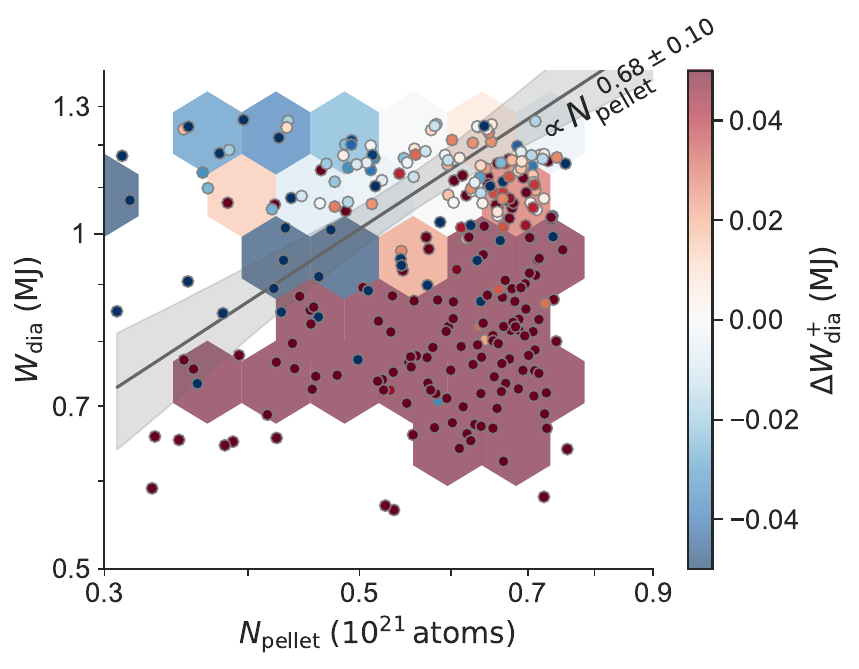}
    \caption{
        Two-dimensional distribution of the measured $\Delta W_\mathrm{dia}^+$ as a function of $W_\mathrm{dia}$ and $N_\mathrm{pellet}$.
        Each marker indicates a pellet-injection event, and its color indicates the associated value of $\Delta W_\mathrm{dia}^+$.
        The colored hexagonal tiles show the median value in each region and thus represent the overall trend in the two-dimensional space.
        The $\Delta W_\mathrm{dia}^+=0$ boundary (the white region between the blue and red regions) indicates the quasi-steady-state value of $W_\mathrm{dia}$ for a given $N_\mathrm{pellet}$.
        The solid line is a power-law fit to this boundary obtained by logistic regression, $W_\mathrm{dia}\propto N_\mathrm{pellet}^{0.68\pm0.10}$.
        Both axes use logarithmic scales.
    }
    \label{fig:heatmap}
\end{figure}

\Fref{fig:heatmap} presents $\Delta W_\mathrm{dia}^+$ as a function of \Wdia and $N_\mathrm{pellet}$, which simultaneously shows \fref{fig:statistics}~(c-1) and (c-2).
Markers denote pellet-injection events, with color indicating $\Delta W_\mathrm{dia}^+$ for each event.
The color of the underlying hexagonal tiles indicates the median value of $\Delta W_\mathrm{dia}^+$ calculated from the events in the region.
In the lower-\Wdia region, many points are red (positive $\Delta W_\mathrm{dia}^+$).
This result indicates that plasmas with low \Wdia can be improved by pellets over the full range of sizes studied.
In the upper-left corner, the cluster of blue points indicates that injecting a small pellet into a high-\Wdia plasma degrades its performance.

The white tiles between the red and blue regions represent the $\Delta W_\mathrm{dia}^+=0$ boundary.
This boundary represents the \textit{quasi-steady-state} condition.
A clear positive slope is evident: as $N_\mathrm{pellet}$ increases, the boundary moves upward.
To quantify the boundary position, we perform logistic regression on the pellet-injection data.
The regression identifies the most probable boundary, expressed in the power-law form $\propto W_\mathrm{dia}^\alpha N_\mathrm{pellet}^\beta$, that predicts the sign of $\Delta W_\mathrm{dia}^+$.
The prediction is shown by the solid line, and the $2\sigma$ uncertainty is indicated by gray shading.
The boundary scales as $W_\mathrm{dia}\propto N_\mathrm{pellet}^{0.68\pm0.10}$, where the uncertainty represents one standard deviation.
This result demonstrates the positive dependence of steady-state \Wdia on $N_\mathrm{pellet}$: larger pellets can sustain plasmas with higher \Wdia.

\section{Discussion\label{sec:discussion}}

Particle fueling involves pellet ablation ($\sim10^{-4}$~s), rapid equilibration ($\lesssim10^{-3}$~s), and subsequent transport on longer time scales.
The deposition radius $r_\mathrm{deposit}$, shown in \fref{fig:profile}~(b) and \fref{fig:statistics}~(a), characterizes the fueling profile after the rapid ablation and equilibration processes.

Once a pellet enters the plasma, the heat flux ablates the solid hydrogen and forms a dense plasmoid in $\approx 10^{-4}$~s.
The pellet lifetime and the penetration length achieved during that lifetime have been studied both theoretically and experimentally.

The widely used neutral-gas-shielding (NGS) model assumes that an expanding ablation cloud shields the solid pellet from the incident plasma heat flux~\cite{Parks1978-wm,Houlberg1988-gd}.
Assuming linear radial profiles of the electron temperature and density, each peaked at the plasma axis and zero at the edge, the penetration length is predicted as~\cite{Baylor1997-oj}
\begin{align}
    \lambda_\mathrm{NGS}/a_0 = C T_\mathrm{e,0}^{-5/9} n_\mathrm{e,0}^{-1/9} N_\mathrm{pellet}^{5/27} v_\mathrm{pellet}^{1/3},
\end{align}
where the central electron temperature $T_\mathrm{e,0}$ is expressed in keV, the central electron density $n_\mathrm{e,0}$ in $\mathrm{m^{-3}}$, the pellet size $N_\mathrm{pellet}$ in atoms, and the pellet velocity $v_\mathrm{pellet}$ in $\mathrm{m/s}$.
Here, $C\approx 0.0026 \,\mathrm{keV^{5/9}\,m^{-2/3}\,s^{1/3}}$ is a constant.
Baylor et al. compiled an experimental pellet database spanning multiple tokamaks, including JET and DIII-D, that exhibits a similar scaling~\cite{Baylor1997-oj}.
Baldzuhn et al. also tested a similar scaling for the W7-AS stellarator~\cite{Baldzuhn2004-vy}.

\begin{derivation}
\begin{quote}
    The original NGS model is written with
    \begin{itemize}
        \item $T_\mathrm{e,0}$ in keV
        \item $n_\mathrm{e,0}$ in $10^{20}\,\mathrm{m^{-3}}$
        \item $N_\mathrm{pellet}$ in $10^{20}$ atoms
        \item $v_\mathrm{pellet}$ in $\mathrm{m/s}$.
    \end{itemize}
    and has the form
    $C = 0.079$.
    Here, we want to use 
    \begin{itemize}
        \item $n_\mathrm{e,0}$ in $\mathrm{m^{-3}}$
        \item $N_\mathrm{pellet}$ in atoms
    \end{itemize} 
    instead, which gives
    $C = 0.079 \times 10^{20/9} \times 10^{-20 \times 5/27} \approx 0.0026$.

    The dimension of $C$ is
    \begin{align*}
        [C] &= \mathrm{keV^{5/9}(m^{-3})^{1/9}(m/s)^{-1/3}} \\
            &= \mathrm{keV^{5/9}m^{-2/3}s^{1/3}}.
    \end{align*}
\end{quote}
\end{derivation}

\begin{figure}[h]
    \centering
    \includegraphics[width=\columnwidth]{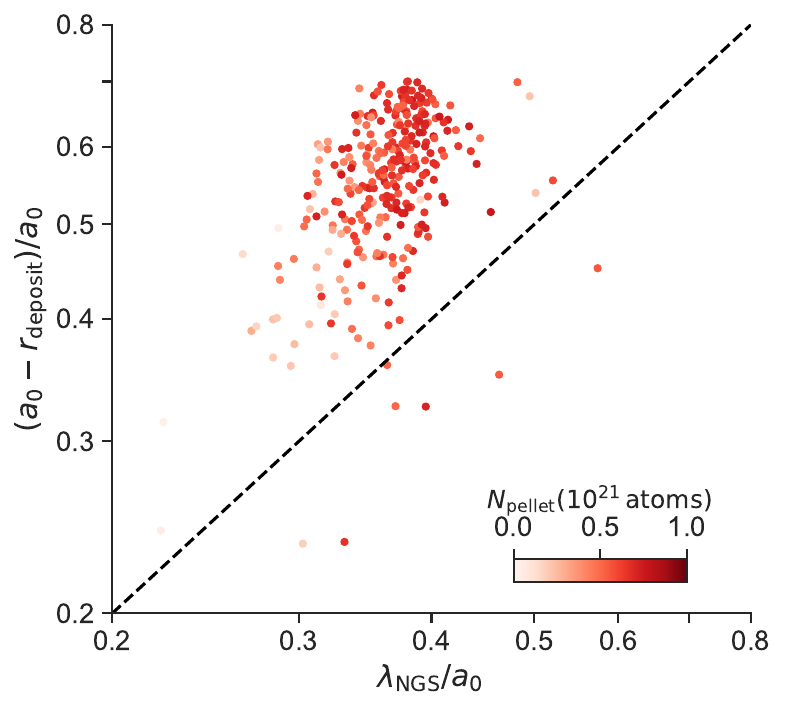}
    \caption{
        Comparison of the measured normalized fueling depth $(a_0-r_\mathrm{deposit})/a_0$ with the NGS-model prediction of the penetration depth $\lambda_\mathrm{NGS}/a_0$.
        Most points lie above the line of equality, indicating greater penetration than predicted by the NGS model and suggesting inward transport of the pellet plasmoid.
    }
    \label{fig:ngs}
\end{figure}

The subsequent equilibration process is less well understood because it involves three-dimensional transport on a short time scale.
Several studies have used high-repetition-rate Thomson scattering to investigate this process.
Funaba et al. measured the spatial distributions of \Te and \Ne during a pellet-injection event in LHD~\cite{Funaba2022-rr}.
Although Thomson scattering samples only a one-dimensional chord, they detected a global decrease in \Te, a local increase in \Ne, and their subsequent equilibration.
At W7-X, Damm et al. used event-synchronized high-repetition-rate Thomson scattering to measure the rapid evolution of \Te and \Ne following hydrogen-pellet injection~\cite{Damm2019-ts}.

Note that Baylor et al. estimated the penetration length from the pellet-light-emission duration and pellet velocity; this estimate characterizes only the ablation process.
Although a fast measurement of the emission duration is unavailable for our experiment, we compare the NGS-model prediction with our values of $r_\mathrm{deposit}$ in \fref{fig:ngs}.
The measured normalized fueling depth $(a_0-r_\mathrm{deposit})/a_0$ is significantly larger than the NGS prediction, which is consistent with the inward transport reported by Damm et al.~\cite{Damm2019-ts}.

The scaling of $(a_0-r_\mathrm{deposit})/a_0$ estimated from the present data suggests a weaker dependence on \Te and a stronger dependence on \Ne than predicted by the NGS model.
In tokamaks, polarization-driven plasmoid drift can produce radial displacement of the ablated pellet material~\cite{Parks2000-vl,Vallhagen2023}.
The pellet-size dependence of the deposition position observed here may therefore be related to plasmoid drift, but a detailed analysis is left for future work.

\section{Conclusion\label{sec:summary}}

We have investigated how pellet size affects the quasi-steady-state performance of W7-X plasmas by analyzing pellet-injection experiments from the OP2.2 and OP2.3 campaigns. Individual injection events show that the plasma response depends strongly on both the pre-injection stored energy and the pellet size. At low $W_\mathrm{dia}$, pellet injection steepens the density profile and increases the stored energy. As $W_\mathrm{dia}$ rises, however, the deposition position moves outward, the pellet-induced change in the density gradient decreases, and the stored-energy gain approaches zero. These trends explain the observed saturation of pellet-fueled plasma performance.

For high-stored-energy plasmas, larger pellets penetrate farther inward, produce a more favorable change in the density gradient, and yield a larger $\Delta W_\mathrm{dia}^+$. 
Logistic regression of the $\Delta W_\mathrm{dia}^+=0$ boundary shows that the quasi-steady-state stored energy scales as $W_\mathrm{dia}\propto N_\mathrm{pellet}^{0.68\pm0.10}$ under the selected conditions. 
Thus, pellet size is not only a fueling parameter but also a control parameter for the sustainable plasma-performance level.

The observed dependencies are qualitatively consistent with pellet-ablation physics: increasing plasma density and temperature shortens the penetration length, whereas increasing pellet size extends it. 
Quantitatively, however, the measured fueling depth is substantially larger than the NGS-model prediction, indicating that rapid inward transport of the ablated pellet cloud contributes to the final fueling profile. 
These results motivate experiments with larger pellets and time-resolved measurements of transport of the pellet ablation cloud to determine whether the quasi-steady-state performance of W7-X can be increased further.

\section*{Appendix: Analysis of the Full Pellet-Injection Dataset}
\setcounter{figure}{0}
\renewcommand{\thefigure}{A\arabic{figure}}
\renewcommand{\theHfigure}{A\arabic{figure}}

The main analysis is restricted to pellet-injection events with steady ECRH power between 5 and 6~MW, no neutral-beam injection, an injection interval longer than 0.3~s, and the high-$\iota$ magnetic configuration. To examine whether the observed pellet-size dependence is specific to these selection criteria, we repeat the analysis here using all available pellet-injection events from OP2.2 and OP2.3. This expanded dataset includes a broader range of heating powers, magnetic configurations, and plasma conditions.

\begin{figure}[t]
    \centering
    \includegraphics[width=\columnwidth]{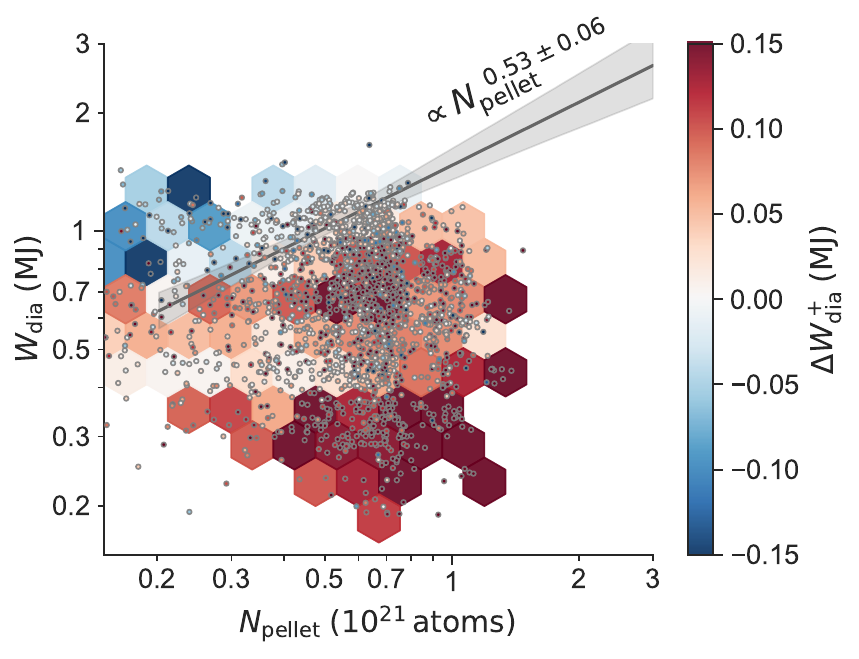}
    \caption{
        Distribution of $\Delta W_\mathrm{dia}^+$ as a function of $W_\mathrm{dia}$ and $N_\mathrm{pellet}$ for all pellet-injection events, including events with different magnetic configurations and heating powers.
        As in \fref{fig:heatmap}, each marker represents one event, the colored hexagonal bins show the median $\Delta W_\mathrm{dia}^+$ in each region, and the solid line denotes the $\Delta W_\mathrm{dia}^+=0$ boundary obtained by logistic regression.
        The fitted boundary scales as $W_\mathrm{dia}\propto N_\mathrm{pellet}^{0.53\pm0.06}$; the shaded region indicates the $2\sigma$ uncertainty.
    }
    \label{fig:heatmap_all}
\end{figure}

\Fref{fig:heatmap_all} shows that the principal trend identified in the controlled subset remains visible in the full dataset. Pellet injection generally increases the stored energy in low-$W_\mathrm{dia}$ plasmas, whereas negative values of $\Delta W_\mathrm{dia}^+$ become more common at high $W_\mathrm{dia}$, particularly for small pellets. The $\Delta W_\mathrm{dia}^+=0$ boundary retains a clear positive slope and scales as $W_\mathrm{dia}\propto N_\mathrm{pellet}^{0.53\pm0.06}$. Thus, the positive dependence of the sustainable stored energy on pellet size is robust across the wider range of operating conditions. The exponent is smaller than the value of $0.68\pm0.10$ obtained from the controlled subset, indicating that variations in heating, magnetic configuration, and plasma state modify the quantitative scaling.

\begin{figure*}[t]
    \centering
    \includegraphics[width=\textwidth]{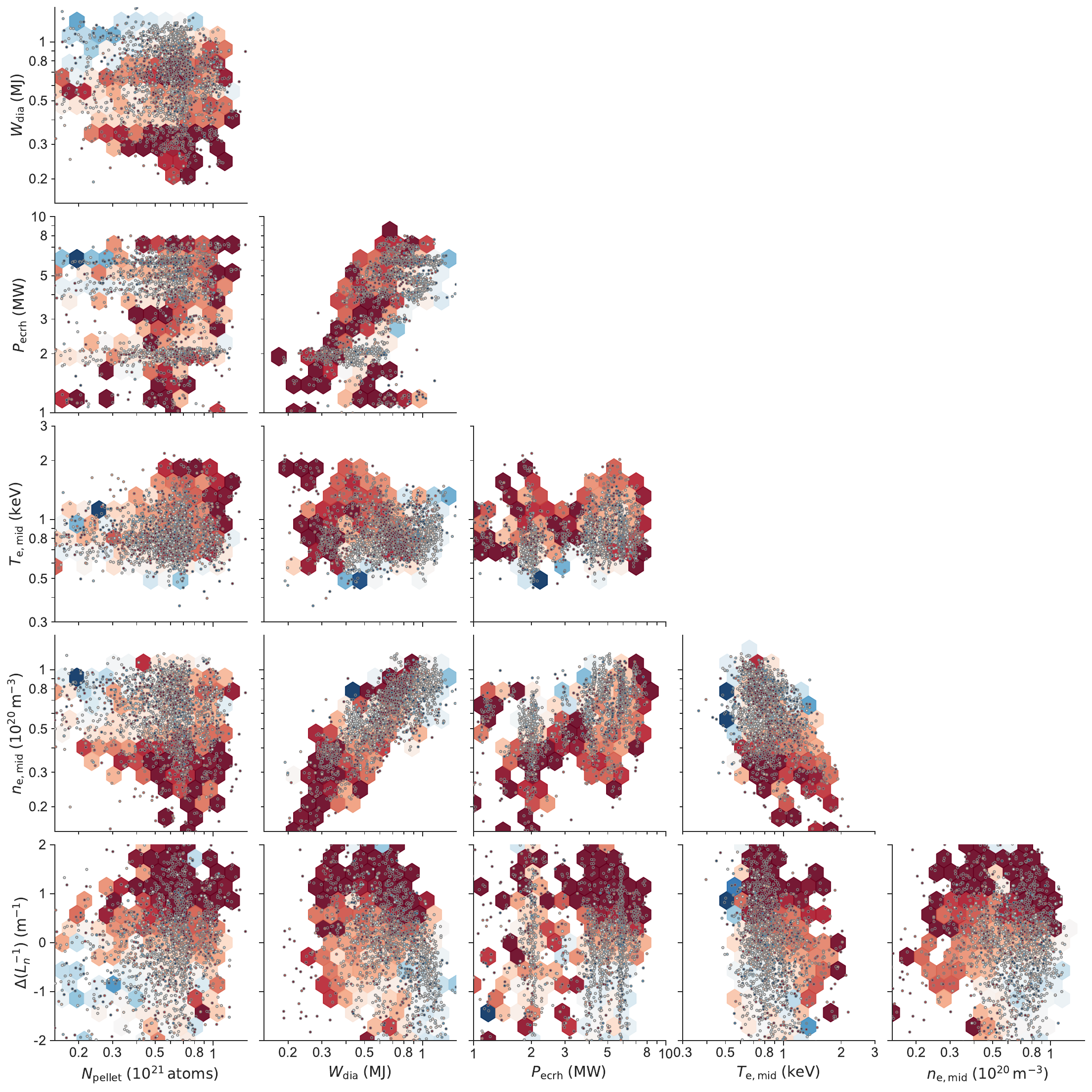}
    \caption{
        Pairwise distributions of pellet size $N_\mathrm{pellet}$, pre-injection stored energy $W_\mathrm{dia}$, ECRH power $P_\mathrm{ecrh}$, midradius electron temperature $T_\mathrm{e,mid}$, midradius electron density $n_\mathrm{e,mid}$, and the pellet-induced change in density gradient $\Delta L_n^{-1}$ for all pellet-injection events.
        Each marker represents one event, and the colored hexagonal bins indicate the median $\Delta W_\mathrm{dia}^+$ within each region.
    }
    \label{fig:multiplot}
\end{figure*}

The pairwise distributions in \fref{fig:multiplot} illustrate why controlling the plasma conditions is important for extracting the pellet-size dependence. The full dataset contains distinct operating regimes, most clearly in $P_\mathrm{ecrh}$, and shows substantial correlations among $W_\mathrm{dia}$, $P_\mathrm{ecrh}$, $T_\mathrm{e,mid}$, and $n_\mathrm{e,mid}$. Consequently, these quantities cannot be treated as independent control variables in the unrestricted dataset. Nevertheless, the color distributions retain the response identified in the main analysis: positive $\Delta W_\mathrm{dia}^+$ is associated with a more favorable pellet-induced change in the density gradient, whereas events with a negative $\Delta L_n^{-1}$ more often exhibit little stored-energy gain or a decrease in $W_\mathrm{dia}$. The persistence of these trends in the full dataset supports the interpretation that pellet size affects plasma performance through the fueling profile and the resulting density-gradient evolution, while the controlled subset provides the more reliable quantitative scaling.

\begin{acknowledgments}
    This work was supported by the U.S. Department of Energy under Contract No.~DE-AC05-00OR22725.
    This work has been carried out within the framework of the EUROfusion Consortium, funded by the European Union via the Euratom Research and Training Programme (Grant Agreement No~101052200~---~EUROfusion). Views and opinions expressed are however those of the author(s) only and do not necessarily reflect those of the European Union or the European Commission. Neither the European Union nor the European Commission can be held responsible for them.
    K.~F. thanks the W7-X diagnostic team, particularly the MHD group, for providing the magnetic-diagnostic data, including \Wdia.
\end{acknowledgments}

\clearpage

\bibliography{refs}



\begin{widetext}
Notice:  This manuscript has been authored by UT-Battelle, LLC, under contract DE-AC05-00OR22725 with the US Department of Energy (DOE). The US government retains and the publisher, by accepting the article for publication, acknowledges that the US government retains a nonexclusive, paid-up, irrevocable, worldwide license to publish or reproduce the published form of this manuscript, or allow others to do so, for US government purposes. DOE will provide public access to these results of federally sponsored research in accordance with the DOE Public Access Plan (\url{http://energy.gov/downloads/doe-public-access-plan}).
\end{widetext}

\end{document}